\documentclass[
  aip,
  jcp,
  amsmath,amssymb,
  reprint,
]{revtex4-1}
\usepackage{graphicx,color}
\usepackage{amsmath}
\usepackage{natbib}
\usepackage{epsfig}
\begin{document}

\title{Instantaneous shear modulus of Yukawa fluids across coupling regimes}

\author{Sergey A. Khrapak}
\affiliation{Institut f\"ur Materialphysik im Weltraum, Deutsches Zentrum f\"ur Luft- und Raumfahrt (DLR), 82234 We{\ss}ling, Germany}

\author{Boris A. Klumov}
\affiliation{Institut f\"ur Materialphysik im Weltraum, Deutsches Zentrum f\"ur Luft- und Raumfahrt (DLR), 82234 We{\ss}ling, Germany}
\affiliation{Joint Institute for High Temperatures, Russian Academy of Sciences, 125412 Moscow, Russia}
\affiliation{Ural Federal University, 620002 Ekaterinburg, Russia}

\date{\today}

\begin{abstract} 
The high frequency (instantaneous) shear modulus of three-dimensional Yukawa systems is evaluated in a wide parameter range, from the very weakly coupled gaseous state to the strongly coupled fluid at the crystallization point (Yukwa melt). This allows us to quantify how shear rigidity develops with increasing coupling and inter-particle correlations. The radial distribution functions (RDFs) needed to calculate the excess shear modulus have been obtained from extensive molecular dynamics (MD) simulations. MD results demonstrate that fluid RDFs appear quasi-universal on the curves parallel to the melting line of a Yukawa solid, in accordance with the isomorph theory of Roskilde-simple systems. This quasi-universality, allows to simplify considerably calculations of quantities involving integrals of the RDF (elastic moduli represent just one relevant example). The calculated reduced shear modulus grows linearly with the coupling parameter at weak coupling and approaches a quasi-constant asymptote at strong coupling. The asymptotic value at strong coupling is in reasonably good agreement with the existing theoretical approximation.  
\end{abstract}

\maketitle

{\it Introduction.} Elastic moduli and related quantities are important characteristics of a material. In this article we calculate the high-frequency  (instantaneous) shear modulus, $G_{\infty}$, which characterizes shear rigidity, for a three-dimensional one-component Yukawa fluid across coupling regimes. The instantaneous shear modulus (as well as bulk modulus) is finite and well defined in fluids, because the fluid response to sudden (high-frequency) perturbations is not much different from that of a solid body.~\cite{ZwanzigJCP1965} Moreover, the properly normalized instantaneous shear modulus is known not to vary much in the dense (strongly-coupled) fluid regime and is numerically close to that of a corresponding (isotropic) solid. In this regime the instantaneous shear modulus becomes an important quantity, which affects and regulates the transverse sound propagation, the instantaneous Poisson's ratio,~\cite{KhrapakPRE2019} the coefficient in the Stokes-Einstein relation,~\cite{KhrapakMolPhys2019} Lindemann melting rule,~\cite{BuchenauPRE2014,KhrapakPRL2019} relaxation time in the shoving model,~\cite{DyrePRE2004,DyreJCP2012} just to mention a few examples.

In this Brief Communication we focus on two main aspects: How the strongly-coupled asymptote of $G_{\infty}$ is approached from the side of disordered weakly coupled gaseous state and how the calculation of $G_{\infty}$ and related quantities can be simplified in the strongly coupled regime by using the ``corresponding states'' approach. 

Our interest to Yukawa fluids is mainly justified by the fact that traditionally Yukawa (screened Coulomb or Debye-H\"uckel) potential is extensively used as a first approximation to model real interactions between charged particles in complex (dusty) plasmas.~\cite{TsytovichUFN1997,FortovUFN,FortovPR,FortovBook,KlumovUFN2011,BonitzRPP2010,ChaudhuriSM2011} The results can also be of some interest in the context of strongly coupled plasmas and colloidal suspensions.~\cite{IvlevBook} In a more general context, the Yukawa potential represents just one particular example of soft repulsive interactions operating in various soft matter systems.    

{\it Formulation}. Yukawa systems represent a collection of point-like charged particles interacting via the pairwise Yukawa (screened Coulomb) potential of the form 
\begin{equation}\label{Yukawa}
\phi(r)=(Q^2/r)\exp(-r/\lambda),
\end{equation}
where $Q$ is the particle charge and $\lambda$ is the screening length. Such a system is fully characterized by the two dimensionless parameters: the coupling parameter $\Gamma=Q^2/aT$ and the screening parameter $\kappa=a/\lambda$, where $a=(4\pi n/3)^{-1/3}$ is the Wigner-Seitz radius,  $T$ is the temperature in energy units ($k_{\rm B}=1$), and $n$ is the density. Conventionally, the systems is referred to as strongly coupled when $\Gamma\gg 1$, that is when the Coulomb interaction energy exceeds considerably the kinetic energy. 

The high-frequency (instantaneous) elastic moduli of simple monoatomic fluids can be related to the pairwise interaction potential $\phi(r)$ and radial distribution function (RDF) $g(r)$. A thorough analysis of the three-dimensional case, with particular emphasis on Lennard-Jones fluids was performed by Zwanzig and Mountain.~\cite{ZwanzigJCP1965} The instantaneous shear modulus can be expressed as~\cite{ZwanzigJCP1965,Schofield1966}
\begin{equation}\label{G1}
G_{\infty}=nT+\frac{2\pi n^2}{15}\int_0^{\infty}dr r^3 g(r)\left[r \phi''(r)+4\phi'(r)\right].
\end{equation}    
The first term above corresponds to kinetic contribution, the second one is the potential (excess) contribution. 

In the ideal gas limit no correlations are present, which corresponds to $g(r) =1$. The excess term vanishes in this regime, because 
\begin{equation}\label{identity}
\Delta G_{\infty}\propto \int_0^{\infty}dr r^3 \left[r \phi''(r)+4\phi'(r)\right]=\left[r^4\phi'(r)\right]_{0}^{\infty}\equiv 0,
\end{equation}     
for potentials that diverge slower than $\propto r^{-4}$ as $r\rightarrow 0$ and decay faster than $\propto r^{-4}$ as $r\rightarrow \infty$. Yukawa interaction belongs to this class. The only contribution to the shear modulus is the kinetic term, $nT$. Substituting this into the expression for Maxwellian shear relaxation time, $\tau=\eta/G_{\infty}$, along with the conventional definition $\tau\sim\ell/v_{\rm T}$ (where $\ell$ is the mean free path and $v_{\rm T}=\sqrt{T/m}$ is the thermal velocity) allows us to reproduce elementary kinetic theory expression for the gaseous shear viscosity~\cite{LL_PhysKin}
\begin{displaymath}
\eta\sim nT\ell/v_{\rm T} =  m v_{\rm T}n\ell,
\end{displaymath}
where $m$ is the particle mass.  Nevertheless, even though the instantaneous shear modulus remains finite at gaseous densities due to the presence of the kinetic term, it is not a very useful quantity in this regime.~\cite{BrazhkinJPCB2018} 

As the coupling increases and the correlations build up in the fluid phase, the excess contribution to the shear modulus becomes progressively more and more important. At sufficiently strong coupling, not too far from the fluid-solid phase transition, transverse (shear) collective excitations can be supported. The specifics of the transverse mode in dense fluids (as compared to solids) is the existence of a minimum threshold wave-number, $k_*$, above which transverse mode exists. This phenomenon, often referred to as the $k$-gap in the transverse dispersion relation constitute an important fundamental research topic across disciplines.~\cite{GoreePRE2012,Trachenko2015,BolmatovPCL2015,YangPRL2017,KhrapakIEEE2018,KhrapakJCP2019,KryuchkovSciRep2019}       
The long-wavelength transverse mode dispersion relation is to a good accuracy described by~\cite{KhrapakJCP2019}
\begin{displaymath}
\omega^2\simeq \frac{G_{\infty}}{mn}k^2-\frac{1}{2\tau^2},
\end{displaymath} 
where $\omega$ is the frequency, $k$ is the wave vector, and  the transverse sound velocity is $c_t^2 =(G_{\infty}/mn)$. In the considered strongly coupled regime, excess contribution to $G_{\infty}$ dominates and properly normalized $G_{\infty}$ and $c_t$ appear quasi  $\Gamma$-independent.~\cite{KalmanPRL2000,KhrapakPoP2016}  In the following we will see how this asymptote is approached from the side of weak coupling.


{\it Structure of Yukawa fluids}. Generally, the RDF $g(r)$ is required as an input for the calculation of elastic moduli, e.g. using Eq.~(\ref{G1}). We have generated a set of RFDs by performing molecular dynamics (MD) simulations using LAMMPS package~\cite{LAMMPS} (a possible alternative approach, without need to perform MD simulations, would be to use an accurate isomorph-based empirically modified hypernetted-chain approach developed recently~\cite{ToliasPoP2019,CastelloPoP2019}). In our MD suimulations, the system of about $6\times 10^4$ particles has been simulated in the Nose-Hoover thermostat and $NVT$ ensemble with periodic boundary conditions. Starting from the crystal state at $\Gamma \sim 3 \Gamma_{\rm m}$, where $\Gamma_{\rm m}$ is the coupling parameter at the fluid-solid phase transition (melting),~\cite{HamaguchiPRE1997} the system has been heated up to $\Gamma \sim 10^{-2} \Gamma_{\rm m}$. Each configuration has been equilibrated during $10^6$ time steps. We have used about $10^3$ of statistically independent configurations to obtain the RDFs. The cutoff radius for the Yukawa interaction potential has been chosen as $r_{\rm cut} \simeq 10/\kappa$.

\begin{figure}
\includegraphics[width=7cm]{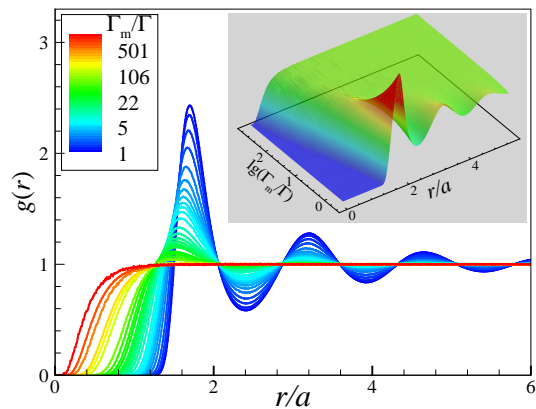}
\caption{Radial distribution function $g(r)$ of Yukawa systems versus the reduced distance $r/a$. Each curve corresponds to a particular value of $\Gamma/\Gamma_{\rm m}$, where $\Gamma_{\rm m}$ is the value of the coupling parameter at the fluid-solid phase transition.~\cite{HamaguchiPRE1997} The range shown spans from $\Gamma/\Gamma_{\rm m}\simeq 10^{-3}$ (weakly coupled disordered gas) to $\Gamma/\Gamma_{\rm m}=1$ (Yukawa fluid at the boundary of the fluid-solid phase transition or Yukawa melt). Inset shows the same data plotted in 3D in coordinates $r/a$ and $\log_{10}(\Gamma_{\rm m}/{\Gamma})$. Note that the color scheme is different from the main figure. The calculation is for the fixed screening parameter $\kappa=4$.}
\label{Fig1}
\end{figure}

\begin{figure}
\includegraphics[width=7cm]{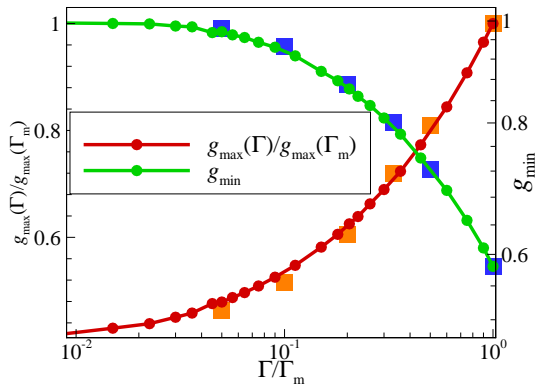}
\caption{The ratio of the amplitude of the first maximum of the RDF to that at the melting point ($g_{\rm max}(\Gamma)/g_{\rm max}(\Gamma_{\rm m})$, left axis) and the amplitude of the first non-zero minimum of the RDF ($g_{\rm min}$, right axis) versus the reduced coupling parameter $\Gamma/\Gamma_{\rm m}$ of Yukawa systems. Circles connected by solid curves correspond to simulations with $\kappa=4$. Squares denote similar data for $\kappa=1$. }
\label{Fig2}
\end{figure}

A representative example is shown in Fig.~\ref{Fig1} where RDFs of a Yukawa fluid with $\kappa=4$ are plotted. RDFs are plotted for various values of the reduced coupling parameter  $\Gamma/\Gamma_{\rm m}$, spanning over very wide range of coupling strength. Note, that $\Gamma_{\rm m}$ is $\kappa$-dependent. A useful approximation of the numerical data for $\Gamma_{\rm m}(\kappa)$ tabulated in Ref.~\onlinecite{HamaguchiPRE1997} is provided by a simple empirical formula~\cite{VaulinaJETP2000}
\begin{displaymath}
\Gamma_{\rm m}(\kappa)\simeq \frac{172 \exp(\alpha\kappa)}{1+\alpha\kappa+\tfrac{1}{2}\alpha^2\kappa^2},
\end{displaymath}
where the constant $\alpha=(4\pi/3)^{1/3}\simeq 1.612$  is just the ratio of the mean interparticle distance $\Delta=n^{-1/3}$ to the Wigner-Seitz radius $a$.  The figure demonstrates how correlations build up as the reduced coupling parameter increases. Two stages can be clearly identified.~\cite{OttPoP2014} First the correlational hole (a spherical cavity where $g(r)\simeq 0$ around a test particle) grows rapidly as the coupling increases. At the second stage, upon further increase in coupling, the radius of the correlational hole saturates, approaching the average interparticle separation $\simeq \Delta$. The shell structure, characterized by the oscillatory behavior of $g(r)$ emerges. In particular, the magnitude of the first peak increases gradually with increasing the coupling strength. At the same time the magnitude of $g(r)$ at the position of the first minimum decreases. These tendencies are further illustrated in Fig.~\ref{Fig2}.         

It has been long known, from the results of Monte Carlo simulations, that details of the interaction potential have relatively little effect on the structure of fluids near the melting temperature, in particular when extreme cases of hard-sphere and Coulomb interactions are excluded from consideration.~\cite{HansenMolPhys1973} Nowadays  this empirical observation is supported by the concept of isomorphs. Isomorphs correspond to curves of constant excess entropy in the thermodynamic phase diagram.~\cite{SchroderJCP2014}  
For systems characterized by strong virial and potential energy correlations (usually referred to as ``Roskilde-simple'' systems), 
structure and dynamics in properly reduced units are invariant along isomorphs to a good approximation.~\cite{DyreJPCB2014,GnanJCP2009} Many simple systems, including the Yukawa case belong to this class.~\cite{VeldhorstPoP2015} Since melting and freezing curves appear as approximate (although not exact) isomorphs,~\cite{PedersenNatCom2016} parallel curves (not too far from the fluid-solid phase transition) should also be approximate isomorphs. This represents justification of using relative coupling strength $\Gamma/\Gamma_{\rm m}$ as a convenient unified state parameter for strongly coupled Yukawa systems. 

It should be noted that other approaches to introduce the effective coupling strength have been discussed in the literature.~\cite{ClerouinPRE2013,OttPoP2014,ClerouinPRL2016,DesbiensPoP2016} It is particularly tempting to use a one-to-one mapping between the structure of Yukawa systems and Coulomb one-component plasma (OCP), because the properties of the latter system are very well know. The properties of the RDF have been successively used for this purpose in Ref.~\onlinecite{OttPoP2014}. However, since OCP represents an extreme limit of soft long-range interaction potentials, the proposed mapping is effective only in the regime of sufficiently weak screening ($\kappa\lesssim 2$).~\cite{OttPoP2014}  Using the relative coupling strength $\Gamma/\Gamma_{\rm m}$ as a mapping criterion allows us to cover a wider range of coupling parameters. Previously, the ratio $\Gamma/\Gamma_{\rm m}$  was often chosen as an adequate relative coupling strength measure in dusty plasmas.~\cite{VaulinaPRL2002,VaulinaPoP2002,FortovPRL2003,OttPRL2009} It was also used to produce useful scalings of transport and thermodynamic properties of Yukawa systems.~\cite{OhtaPoP2000,RosenfeldPRE2000,RosenfeldJPCM2001,VaulinaPRE2002,KhrapakPoP2012,KhrapakPRE2015,KhrapakJCP2015,KhrapakJPCO2018,
CostigliolaJCP2018,KhrapakAIPAdv2018}  

\begin{figure}
\includegraphics[width=6.8cm]{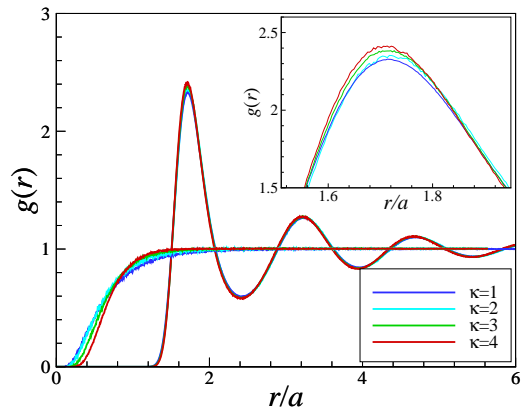}
\caption{Radial distribution functions for the two state points characterized by the same $\Gamma/\Gamma_{\rm m}=1$ (Yukawa melt) and  $\Gamma/\Gamma_{\rm m}=1/200$ (weakly coupled gaseous state) and different screening parameters $\kappa=1,2,3,4$. Inset shows the enlarged portion of strongly coupled RDFs near the first maximum.}
\label{Fig3}
\end{figure}

Figure~\ref{Fig3} shows two sets of RDFs, each calculated for a fixed value of $\Gamma/\Gamma_{\rm m}$, but different values of the screening parameter, $\kappa = 1,2,3,4$. The set with pronounced correlations correspond to  $\Gamma/\Gamma_{\rm m}=1$, that is to strongly coupled Yukawa fluid just near the crystallization point (Yukawa melt). The curves with different $\kappa$ lie almost on top of each other. The small difference is observed in the vicinity of the first maximum: The amplitude of this maximum slightly grows with $\kappa$, as could be expected from our previous experience with inverse-power-law fluids.~\cite{KhrapakSciRep2017} The inset shows the behavior of RDFs near the maximum to illustrate this tendency. When the first maximum of the RDF is normalized by its value at the fluid-solid phase transition, it exhibits a quasi-universal dependence on $\Gamma/\Gamma_{\rm m}$, as documented in Fig.~\ref{Fig2}. Also, the dependence of the magnitude of the first non-zero minimum of the RDF on $\Gamma/\Gamma_{\rm m}$ is quasi-universal, see Fig.~\ref{Fig2}. This can be potentially useful in estimating the relative coupling strength in experiments with complex (dusty) plasma fluids, where RDFs are often easily accessible. At the same time it should be noted that the experimental noise level combined with the relatively smooth dependence of $g(r)$ on  $\Gamma/\Gamma_{\rm m}$ can in many cases hinder the application of this tool.     

The second set of RDFs plotted in Fig.~\ref{Fig3} corresponds to a weakly coupled quasi-gaseous state at $\Gamma/\Gamma_{\rm m}=1/200$. Here the differences between the curves with different $\kappa$ are still small, but observable. This again should be expected, because far from the fluid-solid phase transition the isomorphs are not necessarily parallel to the freezing and melting curves. In fact, deep into weakly coupled gaseous phase, all correlations between the shape of the RDF and melting temperature are lost (see explicit expressions for $g(r)$ below). Nevertheless, even for such small $\Gamma/\Gamma_{\rm m}$ investigated the deviations between RDFs with different $\kappa$ are so tiny that no major effect on the magnitude of integrals involving $g(r)$ should be expected. This is the basis behind the ``corresponding state'' approach used below to reduce the amount of calculations of the shear modulus in the strongly coupled regime.        

{\it Instantaneous shear modulus}. For the Yukawa interaction potential (\ref{Yukawa}) the expression for the instantaneous shear modulus (\ref{G1}) becomes
\begin{equation}\label{Shear_Yukawa}
\Delta G_{\infty}= \frac{m n \omega_{\rm p}^2 a^2}{30} \int_0^{\infty} dx x g(x)e^{-\kappa x}\left(\kappa^2x^2-2\kappa x -2\right),
\end{equation} 
where $x=r/a$ and $\omega_{\rm p}=\sqrt{4\pi Q^2 n/m}$ is the plasma frequency. In the following we will be dealing with the reduced (dimensionless) quantity $\Delta G_{\infty}/m n \omega_{\rm p}^2 a^2$. This is equivalent to expressing the transverse sound velocity in units of $\omega_{\rm p}a$, a common practice in the dusty plasma literature.~\cite{KalmanPRL2000,DonkoJPCM2008,KhrapakPoP2016,KhrapakPoP2016_Relations} An alternative option would be to express the instantaneous shear modulus in units of $nT$ (and, hence, transverse sound velocity in units of $v_{\rm T}$). The relation between the two normalizations is straightforward by virtue of the identity $m n \omega_{\rm p}^2 a^2/nT=3\Gamma$. 

\begin{figure}
\includegraphics[width=7cm]{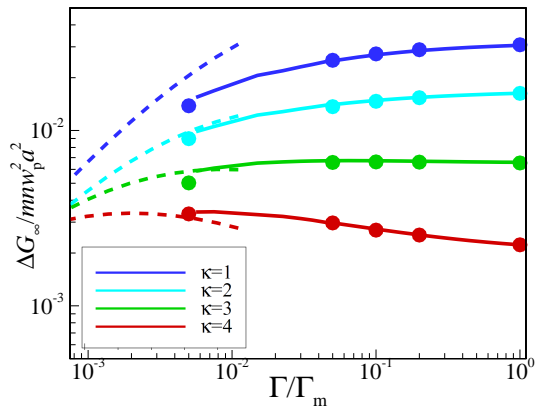}
\caption{Reduced excess instantaneous shear modulus, $\Delta G_{\infty}/m n \omega_{\rm p}^2 a^2$, versus the relative coupling parameter $\Gamma/\Gamma_{\rm m}$ for various screening parameters $\kappa=1,2,3,4$. The symbols correspond to the calculations with ``exact'' RDFs. Solid curves are calculated using ``universal'' set of ($\Gamma/\Gamma_{\rm m}$-dependent) RDFs obtained for $\kappa=4$. Dashed curves are obtained using the Boltzmann approximations for the RDFs in the weakly coupled regime. The transition from the weakly coupled to the strongly coupled regime takes place in the range $10^{-3}\lesssim \Gamma/\Gamma_{\rm m}\lesssim 10^{-2}$. For $\Gamma/\Gamma_{\rm m}\gtrsim 0.1$ the reduced shear modulus approaches a quasi-constant value.}
\label{Fig4}
\end{figure}

The integration in Eq.~(\ref{Shear_Yukawa}) has been performed using the RDFs generated in MD runs and the results are shown in Fig.~\ref{Fig4}. Two approaches have been employed. The direct one is to employ ``exact'' RDFs for each pair of state variables ($\kappa$, $\Gamma$). These results are shown by symbols. The second (approximate) method is what we call here the ``corresponding state'' approach. In this method we use only the set of RDFs calculated for $\kappa=4$ (see Fig.~\ref{Fig1}).  We use an RDF from this set, corresponding to a certain value $\Gamma/\Gamma_{\rm m}$, for other values of $\kappa$ with the same reduced coupling $\Gamma/\Gamma_{\rm m}$. The results from these calculations are shown by the solid curves. We observe that the ``exact'' and approximate approaches demonstrate very good agreement in the strongly coupled regime. The deviations are only observable at the lowest relative coupling  $\Gamma/\Gamma_{\rm m}=1/200$. Thus, there is a wide region where the simple ``corresponding state'' principle can be useful to simplify calculations of various Yukawa fluids properties, which involve integrals over RDFs (elastic moduli represent just one example; other examples include excess energy and pressure, Einstein frequency, frequency moments, etc).         

In the strongly coupled regime ($\Gamma/\Gamma_{\rm m}\gtrsim 0.1$) the instantaneous shear modulus approaches its asymptotic value, characteristic of both fluid and solid. For weak screening this value is approached from below ($\kappa =1$ and 2), while for the highest value investigated ($\kappa=4$) it is approached from above. The existence of a local maximum at relatively weak coupling is unexpected. However, it should be reminded that the reduced (normalized) quantity is plotted. The actual instantaneous shear modulus is expected to increase monotonously on approaching the melting temperature. 

The obtained asymptotic values at strong coupling can be compared with theoretical predictions. Recently, a unified description of elastic moduli of strongly coupled Yukawa systems of different spatial dimensionality has been proposed (main results are expressed in terms of the longitudinal and transverse sound velocities, directly related to elastic moduli).~\cite{KhrapakPoP2019} In this approximation the elastic moduli are related to the internal energy of Yukawa solids using relatively weak sensitivity of the RDFs to the screening parameter at weak screening and the fact that the internal energy is dominated by the static contribution. The excess contribution to the instantaneous shear modulus is then expressed in terms of Madelung constant and its first two derivatives with respect to $\kappa$ (for details see Ref.~\onlinecite{KhrapakPoP2019}). Using the ion sphere model~\cite{RosenfeldJCP1995,KhrapakPoP2014}  as a proxi for the Madelung constant the following expression can be derived~\cite{KhrapakPoP2019}
\begin{equation}\label{Theo}
\frac{\Delta G_{\infty}}{m n \omega_{\rm p}^2 a^2}=\frac{\kappa^4 \left[\left(\kappa^2+3\right) \sinh (\kappa)-3 \kappa \cosh (\kappa)\right]}{45 [\kappa \cosh (\kappa)-\sinh (\kappa)]^3}.
\end{equation}
Comparison between the values obtained from MD-generated RDFs and the theoretical approximation (\ref{Theo}) is provided in Table~\ref{Tab1}. Good agreement is documented. 
 

\begin{table}
\caption{\label{Tab1} Reduced excess contribution to the instantaneous shear modulus of strongly coupled Yukawa systems for different screening parameters $\kappa$. 
}
\begin{ruledtabular}
\begin{tabular}{lcc}
$\kappa$ & MD RDFs [Eq.~ (\ref{Shear_Yukawa})] &  Theory [Eq.~(\ref{Theo})]   \\ \hline
1  & 0.0305 & 0.0319      \\
2  & 0.0163 & 0.0169      \\
3  & 0.0065 & 0.0065      \\
4  & 0.0022 & 0.0020      
\end{tabular}
\end{ruledtabular}
\end{table}   
 
In the limit of vanishing particle density (weakly coupled Yukawa gas), the RDF can be approximated by the corresponding Boltzmann factor, $ g(r)\simeq \exp\left[-\tfrac{\phi(r)}{T}\right]$. Substituting this into equation (\ref{Shear_Yukawa}) results in the dashed curves shown in Fig.~\ref{Fig4}. The transition between weakly coupled and strongly coupled regimes occurs in the region $10^{-3}\lesssim \Gamma/\Gamma_{\rm m}\lesssim 10^{-2}$. This transition is smooth, no special features are observable. Expanding the exponential factor and using the identity (\ref{identity}) we immediately obtain the linear initial increase of the excess shear modulus with $\Gamma$: 
\begin{equation}
\frac{\Delta G_{\infty}}{m n \omega_{\rm p}^2 a^2}\simeq \frac{\Gamma}{24\kappa}.
\end{equation}        
This scaling applies for $\Gamma\ll 1$.       

{\it Conclusion}. We have investigated how shear rigidity is build up when increasing the coupling strength in three-dimensional one-component Yukawa system. The rigidity is characterized here by the high frequency (instantaneous) shear modulus, which can be expressed using the  pairwise interaction potential and the radial distribution function. The latter has been obtained from MD numerical simulations in a wide parameter range across coupling regimes. Simulations support the ``coresponding state'' approach,  in accordance with the isomorph theory of Roskilde-simple systems: The RDFs calculated along the curves parallel to the melting curves in ($\kappa$, $\Gamma$) plane are quasi-universal, at least in the range $1\leq \kappa \leq 4$ and $\Gamma/\Gamma_{\rm m}\gtrsim 0.01$. This allows to considerably simplify the calculation of system properties that involve integrals over the RDF, like the instantaneous shear modulus considered here. The calculated reduced excess shear modulus exhibits the linear increase with $\Gamma$ in the weakly coupled gaseous limit. Then it increases monotonously (for low $\kappa$) or behaves weakly non-monotonously (higher $\kappa$) and approaches the strong coupling asymptotic value. This value is characteristic of strongly coupled fluid and solid phases and is demonstrated to be in good agreement with available theoretical predictions.            

The work leading to this publication was partly supported by the German Academic Exchange Service (DAAD) with funds from the German Aerospace Center (DLR). We would like to thank Hubertus Thomas for reading the manuscript.

%
%
%

\bibliographystyle{aipnum4-1}
\bibliography{Shear_References}

\end{document}